\documentclass[pra, aps, longbibliography, twocolumn, amsmath, amssymb, superscriptaddress,showpacs]{revtex4-1}
\usepackage{graphicx, array, booktabs, color}
\usepackage[colorlinks,linkcolor=blue,anchorcolor=blue,citecolor=blue,urlcolor=blue]{hyperref}

\begin{document}

\title{Simulation of Quantum Computing on Classical Supercomputers}

\author{Ya-Qian Zhao}
\affiliation{State Key Laboratory of High-End Server $\&$ Storage Technology, Jinan 250014, China}
\affiliation{Inspur Electronic Information Industry Co., Ltd, Jinan 250014, China}
\author{Ren-Gang Li}
\affiliation{State Key Laboratory of High-End Server $\&$ Storage Technology, Jinan 250014, China}
\affiliation{Inspur Electronic Information Industry Co., Ltd, Jinan 250014, China}
\author{Jin-Zhe Jiang}
\affiliation{State Key Laboratory of High-End Server $\&$ Storage Technology, Jinan 250014, China}
\affiliation{Inspur Electronic Information Industry Co., Ltd, Jinan 250014, China}
\author{Chen Li}
\affiliation{State Key Laboratory of High-End Server $\&$ Storage Technology, Jinan 250014, China}
\affiliation{Inspur Electronic Information Industry Co., Ltd, Jinan 250014, China}
\author{Hong-Zhen Li}
\affiliation{State Key Laboratory of High-End Server $\&$ Storage Technology, Jinan 250014, China}
\affiliation{Inspur Electronic Information Industry Co., Ltd, Jinan 250014, China}
\author{En-Dong Wang}
\affiliation{State Key Laboratory of High-End Server $\&$ Storage Technology, Jinan 250014, China}
\affiliation{Inspur Electronic Information Industry Co., Ltd, Jinan 250014, China}
\author{Wei-Feng Gong}
\affiliation{State Key Laboratory of High-End Server $\&$ Storage Technology, Jinan 250014, China}
\affiliation{Inspur Electronic Information Industry Co., Ltd, Jinan 250014, China}
\author{Xin Zhang}
 \email{xzphys@gmail.com}
\affiliation{State Key Laboratory of High-End Server $\&$ Storage Technology, Jinan 250014, China}
\affiliation{Inspur Electronic Information Industry Co., Ltd, Jinan 250014, China}
\author{Zhi-Qiang Wei}
\affiliation{Pilot National Laboratory for Marine Science and Technology(Qingdao), Qingdao 266200, China}
\affiliation{College of Information Science and Engineering, Ocean University of China, Qingdao 266100, China}

\date{\today}

\begin{abstract}
Simulation of quantum computing on supercomputers is a significant research topic, which plays a vital role in quantum algorithm verification, error-tolerant verification and other applications. Tensor network contraction based on density matrix is an important single-amplitude simulation strategy, but it is hard to execute on the distributed computing systems currently. In this paper, we dive deep into this problem, and propose a scheme based on cutting edges of undirected graphs. This scheme cuts edges of undirected graphs with large tree width to obtain many undirected subgraphs with small tree width, and these subgraphs contracted on different computing cores. The contraction results of slave cores are summarized in the master node, which is consistent with the original tensor network contraction. Thus, we can simulate the larger scale quantum circuit than single core. Moreover, it's an NP-hard problem to find the global optimum cutting edges, and we propose a search strategy based on a heuristic algorithm to approach it. In order to verify the effectiveness of our scheme, we conduct tests based on QAOA algorithm, and it can simulate 120-qubit 3-regular QAOA algorithm on 4096-core supercomputer, which greatly exceeds the simulation scale on a single core of 100-qubit.

\end{abstract}

\maketitle

\section{introduction}
Quantum computing is a progressive computing mode based upon the principles of quantum entanglement and state superposition, which can bring powerful quantum parallelism, and also bring potential solutions for the problem of insufficient of computing power in the post-Moore era \cite{2000Quantum}. Actually, Feynman has proposed the concept of quantum computing decades ago, which responded the problem of exponential growth of memory requirement on classical computers when simulating quantum systems \cite{1982Simulating}. After decades of development, quantum computing has made great progress in both hardware and algorithm, especially as Google claims to achieve ``quantum supremacy" in 2019, and quantum computing begins to enter the public sight \cite{2019Quantum}. However, quantum computing is still in its infancy overall, and there is still a long way to realize large-scale error-tolerant quantum computer. Considering this background, it is of great significance to build a quantum computing simulation platform based on classical computers, (1) it can provide a reliable verification platform for quantum algorithms and quantum noise \cite{2019On, 2017Quantum, 201864, 2018Massively, 2018Classical}, (2) and can help us understand the boundaries of classical computing and quantum computing \cite{2018A}, to promote the development of quantum computing.

Quantum computing simulation is a relatively new research direction, and it include full amplitude mode and single amplitude mode. The full-amplitude mode needs to store all the amplitudes of eigenstate of quantum states, and the amplitude can be evolved through the quantum gate. The vector dimension of quantum states is $2^{N}$ for N-qubit, the storage requirement increases exponentially with the increase of the qubit number, so it is difficult to simulate a quantum system with more than 45 qubits, even on a large supercomputer. Recently, researchers have made great progress in full amplitude simulation, such as partial amplitude simulation \cite{2020Breaking}, Auxiliary variable method \cite{2018Massively}, MPS and PEPS technologies \cite{2018Validating, PhysRevLett.123.190501}, these new technologies can realize a large-scale simulation of full amplitude mode with more than 45 qubits. Single amplitude mode is a recently developed strategy, which need not to store all the amplitude of the quantum state, and can directly calculate the probability amplitude of the POVM (Positive-Operator Valued Measure) element \cite{2018Classical}. The single amplitude mode can easily simulate quantum supremacy circuits, even exceeding 100 qubits with shallow layers \cite{2018Classical, 2018qTorch}. Firstly, the single amplitude mode maps the quantum circuit to a tensor network, and then contracts it to a rank-0 tensor, which corresponding to the probability amplitude of the POVM element. There are two strategies for single amplitude simulation, one is based on path integration \cite{2018Classical, 2018Simulation}, and the other is based on density matrix currently \cite{2008Quantum}, which make it possible to simulate a 40-layers quantum supremacy circuit \cite{2018Classical}. The strategy based on density matrix is widely used in quantum algorithm simulation and quantum many-body system simulation \cite{2007Efficient, 2007Classical, 2006Classical}, but it progresses slowly nowadays, and cannot be implemented on the distributed computing systems.

In this paper, we study the tensor network contraction algorithm based on density matrix strategy in detail, and propose a parallel strategy based on cutting edges of tensor network. This strategy divides a tensor network with large tree width into several tensor networks with small tree width by cutting edges, and then contract these tensor networks with small tree width on different computing cores, and thus to realize distributed computing. However, finding out the appropriate edges to cut is also a tricky problem, which directly affects the tree width of the generated subgraphs, thus affects the final performance eventually. It is an NP-hard problem to decide the proper edges to cut. Therefore, we propose a search strategy based on heuristic algorithm to help find a solution approaching the global optimum. Therefore, we contribute a distributed computing strategy for tensor network contraction based on density matrix, and successfully simulate a 120-qubit 3-regular QAOA algorithm on a 4096-core supercomputer.

\section{Tensor Network Contraction}
\subsection{Tensor and Tensor Network}

\begin{figure}
\includegraphics[width=7.5cm]{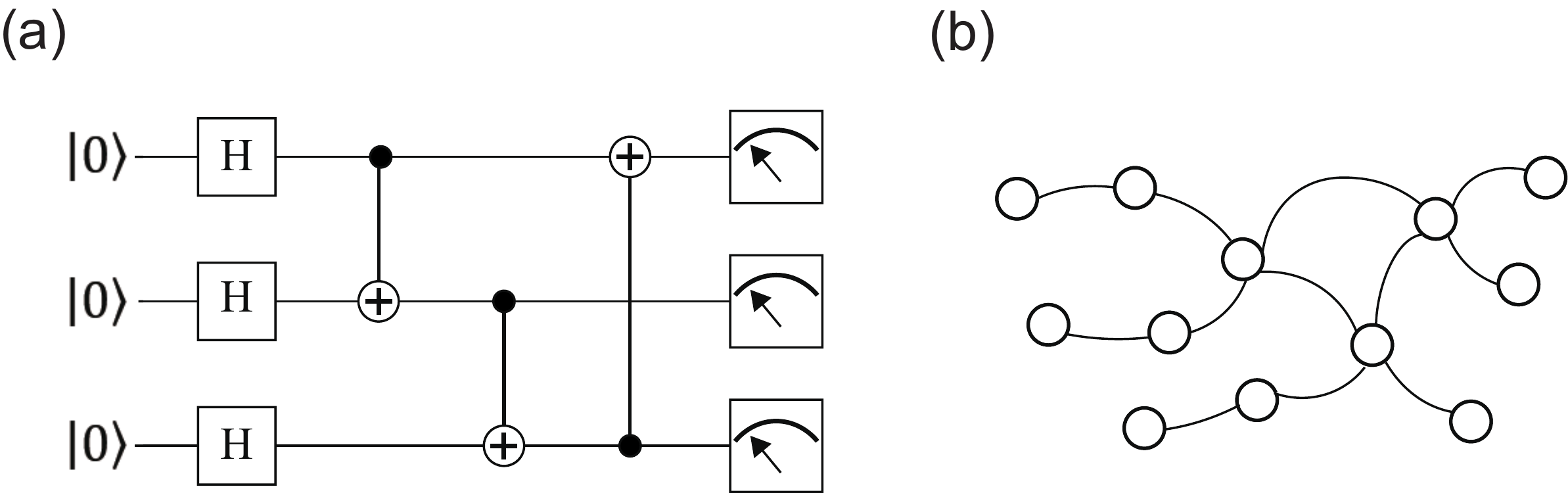}
\caption{(a) The structure of quantum circuit. (b) Undirected graph that corresponds to the quantum circuit. }
\label{fig:fig1}
\end{figure}

Tensor network is the topological connecting of different tensors, which can be represented by an undirected graph generally, and we define it as $G=(V,E)$, where $V$ is the set of vertices and $E$ is the set of edges. As shown in Fig. 1, each quantum circuit corresponds to an undirected graph that represents the corresponding tensor network, and constructing tensor network is the first step to implement tensor network contraction. In the undirected graph, each quantum gate, input, and measurement corresponds to a vertex, and the edge of the quantum circuit corresponds to an edge. Tensor is a data structure with rank and dimension, where the rank means the number of connected edges, which can be indicated by different index (such as $i$, $j$, $k$, $l$, etc.), and the dimension is the number of possible value for each edge. For example, each edge adopt a 4-component density operator with value $\Pi=\{|0\rangle\langle0|, |0\rangle\langle1|, |1\rangle\langle0|, |1\rangle\langle1|\}$, for a k-rank tensor, we can represent it by a one-dimensional array with $4^{k}$ complex numbers.

Markov and Shi gave the method of constructing tensor network in the article \cite{2008Quantum}. For a qubit with an input state $\rho$, its tensor is $T_{\sigma}=\text{tr}(\rho\cdot\sigma)$, where $\sigma\in\Pi$, and the tensor for a single qubit operator is $T_{\sigma,\tau}=\text{tr}(\tau^{\dag}G(\sigma))$. Further, the tensor for a two-qubit operator is $T_{\sigma_{1}, \sigma_{2}, \tau_{1}, \tau_{2}}=\text{tr}((\tau_{1}\otimes\tau_{2})^{\dag}G(\sigma_{1}\otimes\sigma_{2}))$, and the tensor of the measurement operator is $T_{\tau}=\text{tr}(E\cdot\tau)$, where $E$ is the POVM element, and $G$ is the unitary evolution of the operator.

\begin{figure}
\includegraphics[width=7.5cm]{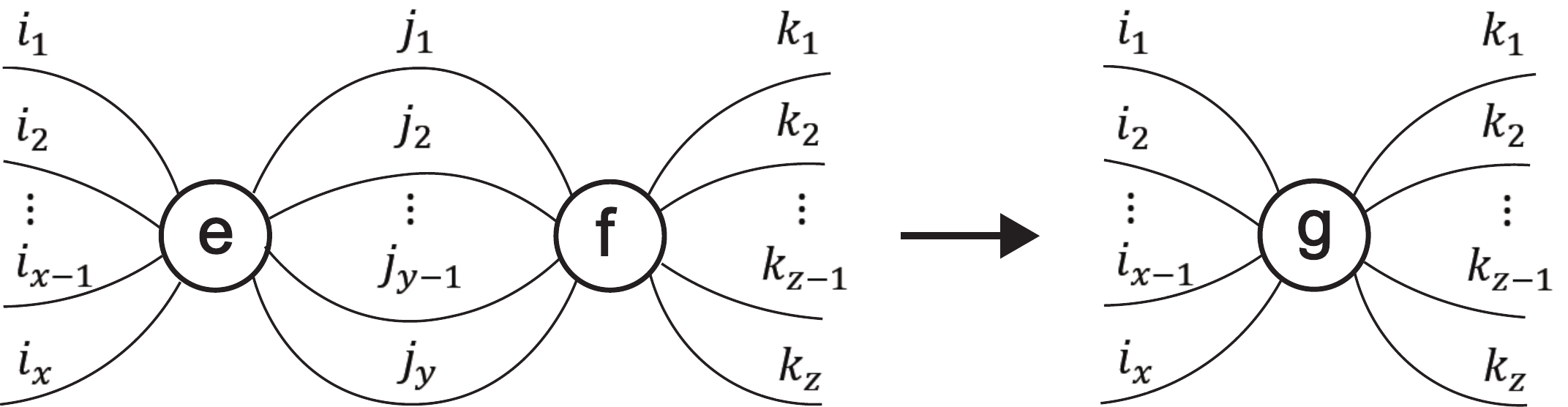}
\caption{A rank $x+y$ tensor and a rank $y+z$ tensor contract to a rank $x+z$ tensor. }
\label{fig:fig2}
\end{figure}

Tensor contraction is a kind of tensor operation, which can contract two connected tensors into one tensor. The two connected tensors have internal edges and open edges, and the tensor contraction is to contract the internal edges and merge two vertices into one. As shown in the Fig. 2, there are two tensors $e$ and $f$, where $e$ is a rank $x+y$ tensor, and $f$ is a rank $y+z$ tensor, and we can get a rank $x+z$ tensor after contraction. The calculation process is as follows,
\begin{eqnarray}\label{ME1}
&&g_{i_{1}, i_{2}, \cdots, i_{x}, k_{1}, k_{2}, \cdots, k_{z}}=\notag\\
&&\sum_{j_{1}, j_{2}, \cdots, j_{y}}e_{i_{1}, i_{2}, \cdots, i_{x}, j_{1}, j_{2}, \cdots, j_{y}}\cdot f_{j_{1}, j_{2}, \cdots, j_{y}, k_{1}, k_{2}, \cdots, k_{z}}
\end{eqnarray}

After the contraction of all the tensors, we can get a rank 0 tensor, which is the probability amplitude of the POVM element.

\subsection{Tree width and tree decomposition}
The maximum memory overhead of the tensor network contraction depends on the tensor of maximum rank in the contraction process. Generally, the maximum rank of the intermediate tensor rise firstly and then fall in the contraction process. For instance, a rank 3+2 tensor and a rank 2+3 tensor will get a rank 3+3 tensor after contraction. Meanwhile, the maximum rank of the intermediate tensor in the process of contraction is related to the order of tensor contraction, where each contraction order corresponds to a tree decomposition of the graph, and the optimal contraction order is the tree decomposition with the smallest tree width \cite{2008Quantum}.

We define $G=(V,E)$ as a undirected graph, the subset of the vertexes of graph $G$ composes a bag denoted by $B_{i}$. The tree decomposition of graph $G$ is a tree $\mathcal{T}$ which is composed of bag $B_{i}$, so the tree decomposition can be expressed as a mapping from the vertex $V(G)$ of the graph $G$ to the $B_{i}$. The following conditions need to be satisfied.

(1) $\cup_{i\in V(\mathcal{T})}B_{i}=V(G)$, each vertex in the graph $G$ can be found in the set of vertex of bag $B_{i}$.

(2) $\forall\{u, v\}\in E(G)$, $\exists i\in V(\mathcal{T})$, satisfy $\{u, v\}\in B_{i}$. That is, the two nodes of an edge in the graph $G$ are included in a $B_{i}$ simultaneously.

(3) All bags that containing the same vertex must be connected in $\mathcal{T}$.

For the tree decomposition $\mathcal{T}$, its width is defined as $\max(|B_{v\in V(\mathcal{T})}|-1)$. The tree decomposition of a graph $G$ is not unique, and the tree width of graph $G$ refers to the minimum width of all possible tree decompositions, denote as $\text{tw}(G)$. It is an NP-hard problem to calculate the tree width and the corresponding tree decomposition of a graph $G$. Fortunately, open source software can be used in actual calculations, such as QuickBB \cite{2004A}. Actually, the time overhead of tensor network contraction is also related to the tree width \cite{2008Quantum}.

\section{Our Algorithm}
\subsection{Edge Cutting of Tensor Network}
In this paper, we propose a tensor network contraction algorithm for distributed computing. Different from the vertex elimination strategy \cite{2018Classical}, we propose a cutting edge strategy. In our tensor network, each edge has four different value, $|0\rangle\langle0|$, $|0\rangle\langle1|$, $|1\rangle\langle0|$ and $|1\rangle\langle1|$. As shown in the Fig. 3, when we cut off one edge, and four subgraphs with different initializations are generated, and then we sum all the contraction results of subgraphs, which accords with that of the original graph contraction. The theoretical calculation is as follows,
\begin{eqnarray}\label{ME1}
p&=&\sum_{\cdots, i, j, k, l, o, p, q, \cdots}\cdots T_{i, j, k, l}^{m}T_{i, o, p, q}^{n}\cdots \notag\\
&=&\sum_{\cdots, j, k, l, o, p, q, \cdots}\cdots T_{|0\rangle\langle0|, j, k, l}^{m}T_{|0\rangle\langle0|, o, p, q}^{n}\cdots
\notag\\
&+&\sum_{\cdots, j, k, l, o, p, q, \cdots}\cdots T_{|0\rangle\langle1|, j, k, l}^{m}T_{|0\rangle\langle1|, o, p, q}^{n}\cdots
\notag\\
&+&\sum_{\cdots, j, k, l, o, p, q, \cdots}\cdots T_{|1\rangle\langle0|, j, k, l}^{m}T_{|1\rangle\langle0|, o, p, q}^{n}\cdots
\notag\\
&+&\sum_{\cdots, j, k, l, o, p, q, \cdots}\cdots T_{|1\rangle\langle1|, j, k, l}^{m}T_{|1\rangle\langle1|, o, p, q}^{n}\cdots
\end{eqnarray}

\begin{figure}
\includegraphics[width=7.5cm]{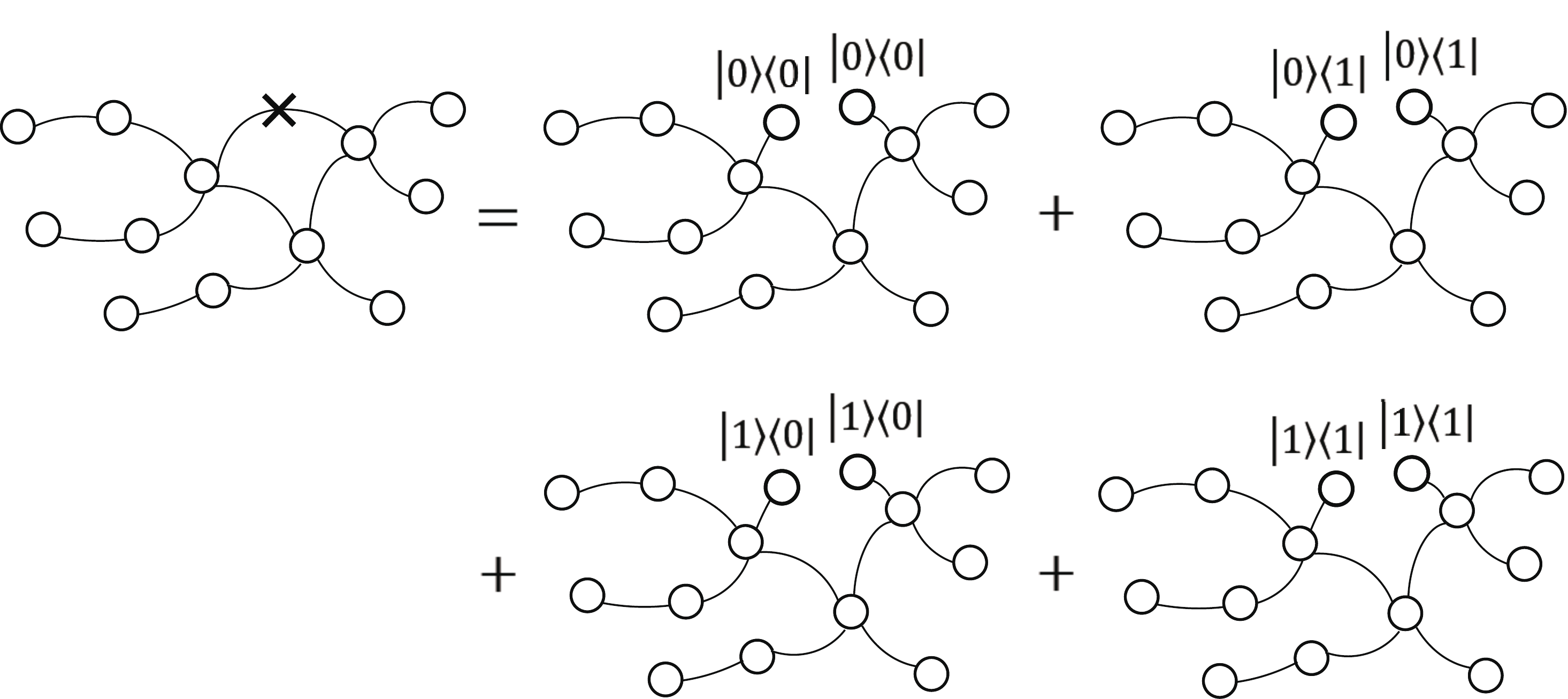}
\caption{The edge cutting of tensor network is to cut off an edge, and add two same additional vertices. Since each edge has four indices, then four subgraphs with different initial values are generated. The subgraphs has a smaller tree width compared with the original graph, so the time complexity and the space complexity of the algorithm are reduced. Each subgraph is contracted on a different computing core, and thus to realize distributed computing. }
\label{fig:fig3}
\end{figure}

We can calculate the contraction of different subgraphs on different computing cores to achieve the distributed tensor network contraction. This method works, because the subgraph after cutting edge has a smaller tree width. It means that the contraction of a subgraph has smaller memory overhead and lower time complexity of algorithm. As shown in Fig. 3, the tree width of original graph is two, and the tree width of subgraph after edge cutting is only one. Actually, we can cut more than one edge, and with more edges to be cut, more subgraphs will be generated, and the tree width of the subgraph become smaller. If m edges are cut, the number of subgraphs is $4^{m}$. If the number of subgraphs is less than that of computing cores, we can directly contract the subgraphs on different cores. But when the number of subgraphs is more than that of computing cores, we need execute the serial calculations for multiple times on the computing cores for all the subgraphs, and the number of the serial times is $\lceil \emph{number of subgraphs} / \emph{number computing cores}\rceil$.

There are several benefits of our method. The main progress is that we only need collect the contraction results of the child processes (The communication data is only a complex value). The communication between child processes is needless, so the communication is not bottleneck of our method. Moreover, the topological structures of subgraphs are completely identical, so the calculation time is also same, and there is no performance waste from the load imbalance.
\begin{table}[htb]
\renewcommand{\arraystretch}{1.5},
\begin{tabular}{m{8.4cm}}
\toprule[1pt]
\textbf{Algorithm 1}. The optimal set searching for edge cutting \\
\midrule[0.5pt]
Initialize iteration counter $t=0$.\\
Set the maximum number of iterations $T$.\\
Initialize population $P$ (Set the population have $N$ individuals, every individual is the set of edge to cut with $M$ elements).\\
\textbf{Repeat:}\\
\hspace{1.3em}\textbf{Step 1:} Individual evaluation. Call QuickBB to calculate the tree width of $N$ individuals of the population $P$, and sort the results.\\
\hspace{1.3em}\textbf{Step 2:} Crossover. Cross every two adjacent individuals except the individual with minimum tree width. Our Crossover method exchanges the $\lfloor M \rfloor$ element of two individuals in the end. If there are duplicate elements in the set after crossover, then randomly generate an element that does not exist in the set to replace one duplicate element.\\
\hspace{1.3em}\textbf{Step 3:} Mutation operation. Randomly select an individual except the optimal individual in the population, then select an edge randomly, and replace it by another edge that does not exist in the individual.\\
\hspace{1.3em}\textbf{Step 4:} Iteration counter increases. $t=t+1$.\\
\textbf{Until} $t>T$.\\
\hspace{1.3em}\textbf{Step 5:} Evaluate individuals in the population and return the best one.\\
\bottomrule[1pt]
\end{tabular}
\end{table}

\subsection{Edge Searching by Genetic Algorithm}
The tree widths of the subgraphs are disparate when cutting different edges, so it's very important to find the set of optimal cutting sequence, which can dramatically improve the performance of the algorithm. It is an NP-hard problem to find the optimal set, so it is impossible to find the solution in a limited time when the graph is very large. In this paper, we propose a strategy for finding the optimal set based on a genetic algorithm, which was shown in Algorithm 1.

Through the above steps, we can search the optimal set of edges to cut approximately, then we can simulate the quantum algorithm on the distributed computing system by the tensor network contraction based on density matric, which have been descripted in section \uppercase\expandafter{\romannumeral3}. A.

\section{Result and Analysis}
\subsection{Experimental Results}
Our test platform is a high-performance scientific computing and simulation platform, which belongs to Pilot National Laboratory for Marine Science and Technology (Qingdao). The platform includes 940 computing nodes, and each node is configured with two Intel Xeon E5-2690v3 or E5-2690v4 processors, so the platform contains 20520 computing cores totally. We implement our simulator by C++ programming language, and use MPI for the communication between processes.

Quantum approximate optimization algorithm (QAOA) is a kind of low-depth quantum circuits. It is used for optimization problem, and is usually utilized to demonstrate quantum supremacy \cite{2014A, 2017Practical, 2016Quantum}. We evaluate our method with 3-regular QAOA, the same as \cite{2018qTorch}. We generate 3-regular graph with a given vertex set, place edges randomly, and abandon the disconnected graphs. In this paper, we only test the single contraction of each graph.

\begin{figure}
\includegraphics[width=7.5cm]{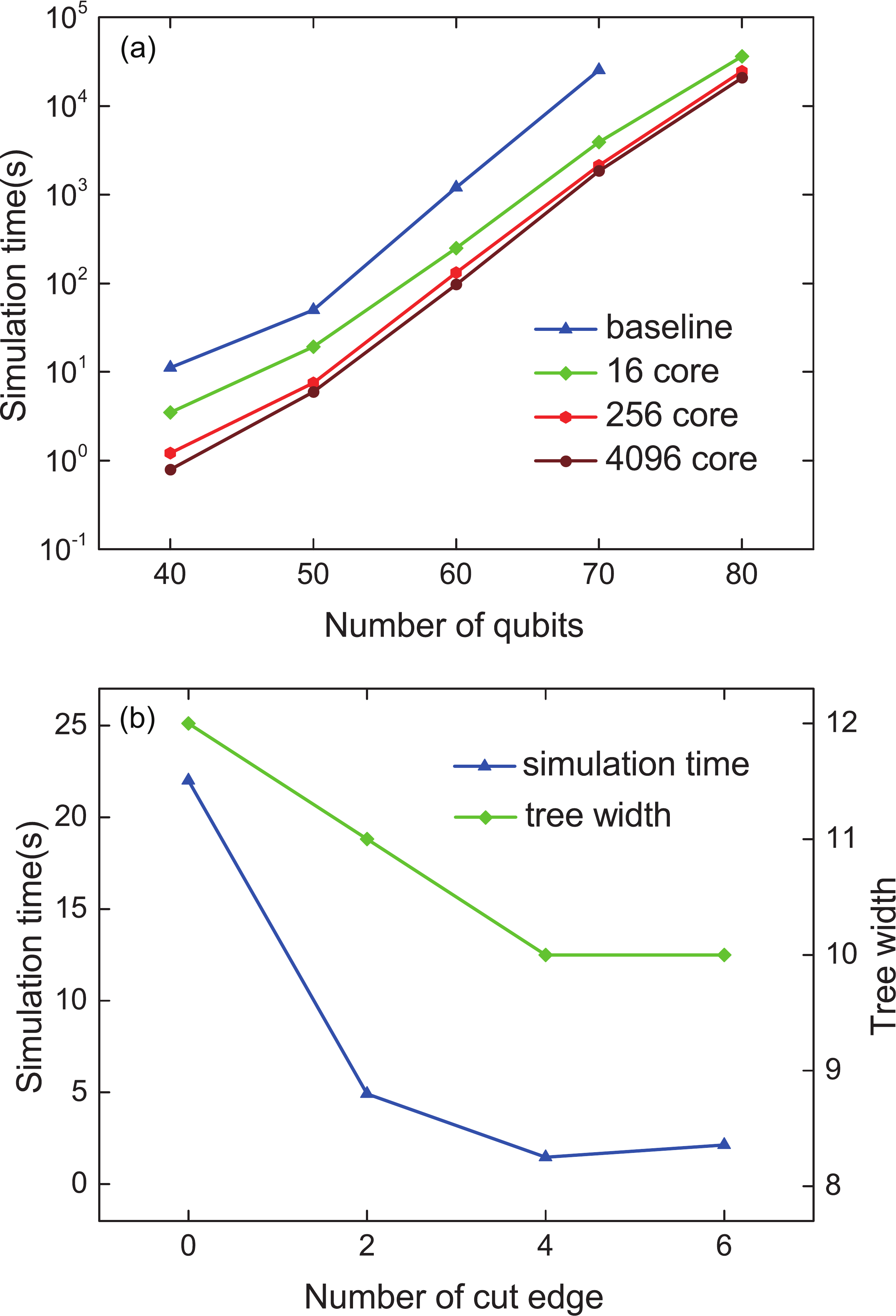}
\caption{(a) Simulation time of QAOA with different qubit numbers, the result is the average of 50 samples. (b) Simulation time and tree width with different numbers of cut edges for a 50-qubit sample. }
\label{fig:fig4}
\end{figure}

In order to verify the effectiveness of our scheme, we test the simulation time of QAOA algorithms with different numbers of vertices on different computing cores. We randomly generate 50 samples for each number of vertices, and calculate the average simulation time, this is same as \cite{2018qTorch}. In our heuristic algorithm, the maximum number of iterations is set as 4, population size is 11.

As shown in Fig. 4(a), we call up to 4096 computing cores. We verify the simulation time of different vertices (qubits) under different computing cores, and find that our algorithm will get better performance with more computing cores. The baseline is showing the performance of qtorch \cite{2018qTorch}, and it is consistent with our simulation performance of single-core. Specifically, each core can contract a tensor network with maximum tree width of 17 in our experiment. In the case of single-core simulation, the tree width of some samples of the 80-qubit exceeds 17, so that we cannot conduct this large-scale simulations, just like the baseline in Fig. 4(a). But if 2 edges are cut off, the tree width is reduced to less than 17, and the simulations with 16 or more cores can be successful. As shown in Fig. 4(b), we select one of the 50-qubit samples, and the lines show that the simulation time and tree width generally decrease as the number of cut edges increases. Actually, the simulation time of different networks with the same tree width may also be different.

In article \cite{2018qTorch}, Fried et al. claimed that they can simulate some samples of 100 qubits for the 3-regular QAOA, which is the largest scale simulation of QAOA by tensor network contraction with density matrix method so far. But in the present paper, we successfully obtain some samples of 120 qubits on a 4096 core supercomputer. Based on the tree width tests of 120 qubit with different numbers of cut edges, we find some samples can get 17 tree width with 8 cut edges, which can fit our server's calculating ability, and convince that we can execute the contraction. Fig. 5 shows the tree width behavior with different cut edges of one sample. We execute the contraction of $4^{8}$ subgraphs on 4096 computing cores, thus we need to serially execute 16 times, and finally we complete the simulation within 382384s.
\begin{figure}
\includegraphics[width=7.5cm]{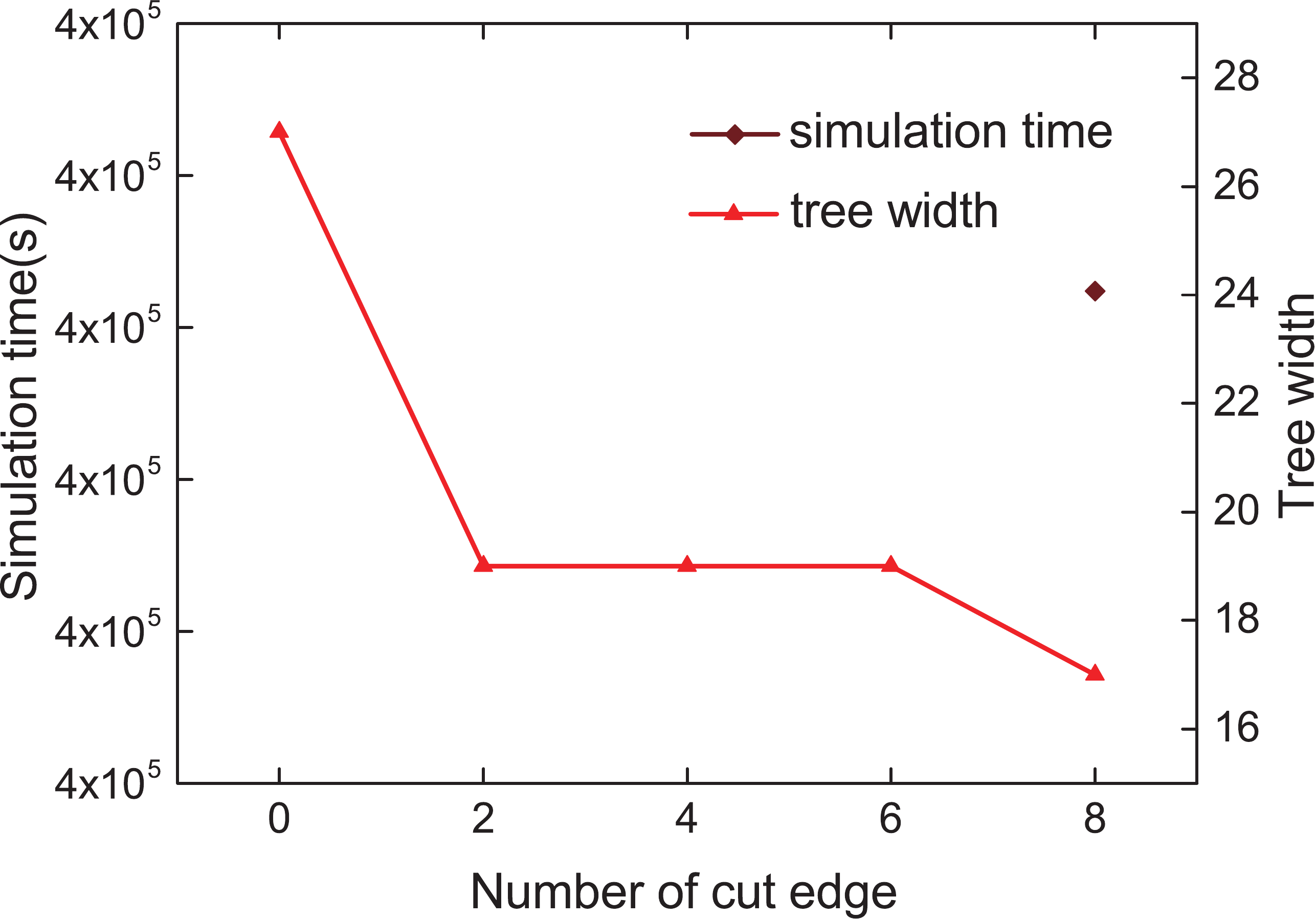}
\caption{Simulation time and tree width with different numbers of cut edges for a 120-qubit sample. }
\label{fig:fig5}
\end{figure}

Actually, we also test the random quantum circuit \cite{2018Characterizing} and Shor algorithm \cite{1997ShorPolynomial}, but the results are disappointing. On a 4096 core supercomputing system, we can only simulate a dozen qubits of random quantum circuit with 40 layers, and two dozen qubits of Shor algorithm.

\subsection{Analysis}
Our scheme is better for shallow and sparse networks, but relatively ineffective for deep and complex networks. Overall, our scheme has the following advantages. (1) It can convert memory overhead into time overhead by controlling the number of cut edges, which can make the balance of the algorithm complexity of space and complexity of time. Therefore, our scheme can simulate any scale quantum algorithm in enough time. (2) The communication overhead between computing nodes is extremely small, and the main process only needs to collect the contraction results of sub-processes, and communication is not a bottleneck at all. (3) The load of the computing core is exactly identical, avoiding performance loss caused by waiting.

\section{Conclusions}
The present paper contributes a distributed computing scheme for the tensor network contraction based on density matrix, and the effectiveness has been verified on classical supercomputing system. We can successfully simulate some samples of 120 qubits 3-regular QAOA on a 4096 cores supercomputer, which dramatically exceeds the simulation scale on a single core computer, and the proposed approach can greatly promote the progress of the scheme of tensor network contraction based on density matrix. Moreover, our scheme can be implemented on very large-scale clusters, and we can perform fault tolerance at the algorithm level, we only need recalculate the missing subgraph when some nodes are down or disconnection.

\section*{Acknowledgments}
The authors thank G. Tian, X. Li, R, Li and R. Zhang for discussions. The authors also acknowledge the Pilot National Laboratory for Marine Science and Technology for the computational resources. This work was supported by National Key R$\&$D Program of China under grant No. 2017YFB1001700, and Natural Science Foundation of Shandong Province of China under grant No. ZR2018BF011.

\end{document}